\documentclass[10pt,conference]{IEEEtran}

\usepackage{hyperref}
\usepackage{cite}
\usepackage{textcomp}
\usepackage{amsmath,amssymb,amsfonts}
\usepackage{xspace}
\usepackage{booktabs}
\usepackage{multirow}
\usepackage{xcolor}
\usepackage{graphicx}
\usepackage{fontawesome}
\newcommand{\approach}{{\sc RepeatNPR}\xspace}
\def\BibTeX{{\rm B\kern-.05em{\sc i\kern-.025em b}\kern-.08em
    T\kern-.1667em\lower.7ex\hbox{E}\kern-.125emX}}
\begin{document}

\title{Neural Program Repair with Program Dependence Analysis and Effective Filter Mechanism}

\author{\IEEEauthorblockN{Yuwei Zhang, Ge Li, Zhi Jin}
\IEEEauthorblockA{
Key Laboratory of High Confidence Software Technologies \\
(Peking University), Ministry of Education\\
Beijing, China \\
\{yuweizhang, lige, zhijin\}@pku.edu.cn}
\and
\IEEEauthorblockN{Ying Xing}
\IEEEauthorblockA{
School of Artificial Intelligence, \\
Beijing University of Posts and Telecommunications\\
Beijing, China \\
xingying@bupt.edu.cn}
}

\maketitle

\begin{abstract}
  Automated program repair is a crucial task for improving the efficiency of software developers. Recently, neural-based techniques have demonstrated significant promise in generating correct patches for buggy code snippets. However, most existing approaches arbitrarily treat the buggy context without any analysis to capture the semantic relationship between the buggy statement and its context. Additionally, we observe that existing neural models may output an unaltered patch consistent with the input buggy code snippet, which fails to be the correct human-written one for fixing the given bug. To address the aforementioned limitations, we present in this paper a novel neural program repair framework called \approach, which adapts the general pre-trained language model for fixing single-line Java bugs. We make the first attempt to use program slicing to extract contextual information directly related to the given buggy statement as repair ingredients from the corresponding program dependence graph and eliminate unaltered patches using an intuitive but effective filter mechanism. We demonstrate the effectiveness of \approach on five benchmarks when compared with state-of-the-art baselines.
\end{abstract}

\section{Introduction}
\label{int}

Program debugging is known to be an extremely labor-intensive and time-consuming task in the software development and maintenance process \cite{jorgensen2007systematic}. To assist software developers in relieving manual debugging efforts, automated program repair (APR), which aims at repairing defective programs without human intervention, has emerged as an important research area in both the software engineering (SE) and artificial intelligence (AI) communities. Over the past decade, a variety of APR approaches \cite{monperrus2018automatic,goues2019automated,gazzola2019automatic,gao2022program} have been proposed to automatically generate patches for bug fixing. Among all traditional approaches, template-based APR \cite{liu2019tbar,koyuncu2020fixminer}, which utilizes predefined patterns to transform buggy code snippets into fixed ones, is often regarded as state-of-the-art \cite{xia2022less,xia2022practical}. However, existing template-based approaches mainly design patterns for specific bug types in a given programming language, which in practice necessitate professional domain knowledge to craft. Therefore, the manually designed templates are challenging to apply to unknown bug types or different programming languages, limiting the effectiveness of APR.

Assisted by the powerful representation ability of deep learning (DL) models, a recent trend is to leverage neural-based techniques to automatically learn intricate relationships between buggy and fixed code snippets from massive source code corpus. Based on the naturalness hypothesis of software code \cite{hindle2016naturalness}, the neural program repair (NPR) approaches intuitively formulate such a problem as the neural machine translation (NMT) task that translates defective programs into correct versions \cite{tufano2019empirical,ding2020patching}. Generally, existing NMT-based NPR systems adopt an encoder-decoder architecture \cite{vaswani2017attention}, where the encoder extracts the hidden status of buggy code snippets with their surrounding context, and the decoder generates fixed code snippets based on the encoder's hidden status. Compared with the previous APR approaches, the advantage of NPR approaches lies in being less dependent on professional domain knowledge and additional artifacts (e.g., test suites). Consequently, researchers are concentrating on advanced NPR approaches, which have shown great potential for automatically generating patches using DL-based techniques.

Despite achieving remarkable performance improvements, existing NPR approaches still have some limitations. First, most NPR models do not fully exploit the contextual information of buggy code snippets. It is vital to delineate what information in the source code should be included in the model input as repair ingredients. However, previous NPR models usually treat the notion of context pertaining to a buggy statement within the corresponding source code in an arbitrary fashion, which neither captures the semantic relationship between the buggy statement and its context nor encompasses essential ingredients for bug fixing. For instance, some NPR models merely take the buggy statement \cite{hata2018learning} as input and ignore context. Others utilize the enclosing buggy class \cite{chen2021sequencer} or the enclosing buggy method \cite{tufano2019empirical,lutellier2020coconut,li2022dear} as context. For software developers, the context of a buggy statement plays a significant role in understanding the root cause of a bug and determining a potential bug-fixing suggestion. Likewise, for the DL-based NPR models, too much information in the input may introduce noise that decreases the repair performance of the model, and too little information may cause relevant repair ingredient loss or overfitting issues. Thus, collecting contextual information related to the buggy code snippets as model input may be more effective in generating correct patches.

\begin{figure}[htbp]
  \centering
  \includegraphics[width=0.47\textwidth]{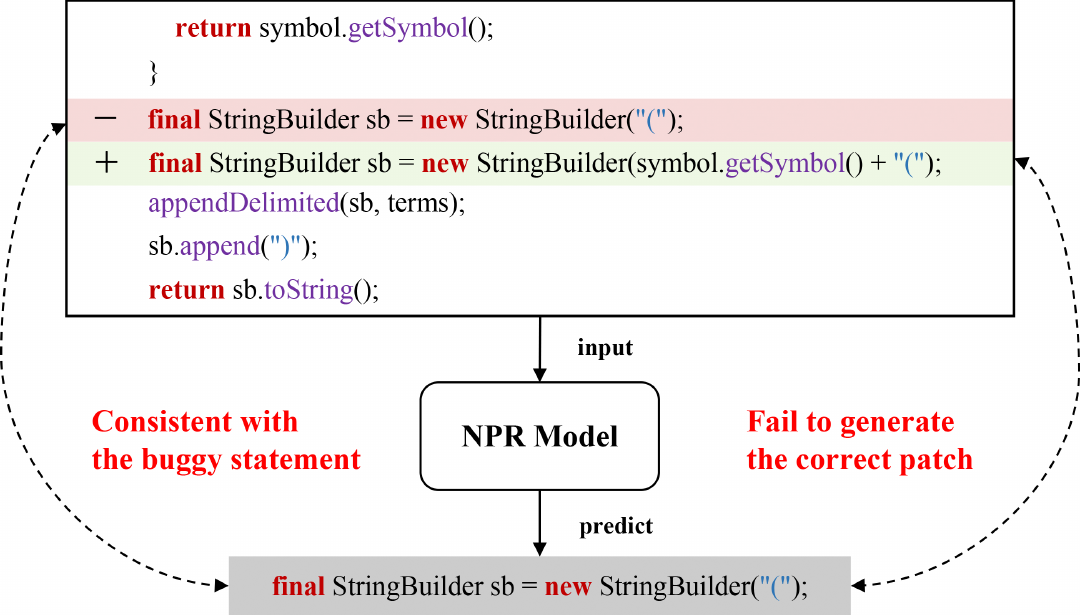}
  \caption{An example of the unaltered patching phenomenon from a real-world open-source software project in GitHub.}
  \label{example}
\end{figure}

Another drawback of current NPR approaches is that the prediction results of many NMT-based NPR models may be consistent with the input buggy code snippets, which we call unaltered patches. The input buggy code snippets frequently share a highly similar vocabulary with the fixed ones, which may cause the model based on NMT architecture to output the same results as the input. Figure~\ref{example} shows an example of a buggy code snippet from an open-source GitHub project\footnotemark[1]. The buggy statement (colored in {\color{red}red}) with its context is regarded as the input for the corresponding NPR model, while the model prediction result (colored in {\color{gray}gray}) is identical to the input buggy statement. Such an unaltered patch fails to be the correct human-written one (colored in {\color{green}green}) for the given bug. This phenomenon is unsupervised because it can determine whether a model-generated patch is correct by comparing it with the input buggy statement without knowing the ground truth. Intuitively, when applying a different NPR model to the unaltered patches, there is a considerable possibility of fixing a portion of them and thus further refining the APR process.

\footnotetext[1]{https://github.com/alpha-asp/Alpha/commit/11c4adc}

In summary, an effective NPR strategy should include adequate contextual information as repair ingredients for patch generation, while filtering out the generated unaltered patches for further processing. To that end, we present an ensemble NPR framework called \approach, which adopts p\textbf{R}ogram d\textbf{EPE}ndence \textbf{A}nalysis with a fil\textbf{T}er mechanism, in this paper to address the limitations of existing approaches. Recently, large language models pre-trained on massive code corpus using code-aware objectives have yielded state-of-the-art results on a vast number of code-related SE tasks. Inspired by this, the proposed framework adapts the general pre-trained language model CodeT5 \cite{wang2021codet5} for fixing bugs of single-line type, that is, we adopt CodeT5 as the starting point to train \approach and fine-tune it for the NPR task.

Specifically, we first employ program slicing based on the def-use analysis \cite{weiser1981program} to extract intra-procedural contextual information in the form of source code token sequences directly related to a buggy statement as its repair ingredients from the program dependence graph (PDG) \cite{ferrante1987program,horwitz1992program}, which explicitly represents both data and control flow dependencies of the corresponding buggy method. In addition, we collect global context according to the ingredients within the intra-procedural contextual information. The slicing-based contextual information is in line with the process of program debugging by software developers. They commonly start by examining all the ingredients (e.g., variables and method invocations) in the buggy statement and then look through the code to find out where those ingredients are defined, used, and modified for understanding the bug \cite{bohme2017bug}. Thus, such relevant contextual information is helpful for the DL-based models to better reason about the bug-fixing process. To tackle the second limitation, we propose an ensemble approach that utilizes an intuitive filter mechanism to combine different NPR models for bug fixing. The filter mechanism directly filters out the unaltered patches by comparing the former NPR model's predicted results with the input buggy code snippets and continues to input the corresponding buggy code snippets into the next NPR model for processing. Just as the cross-checking form in the code review scenario will find more bugs than self-reviewing, an ensemble of different NPR models should also obtain better repair performance than using a single NPR model.

In this paper, the proposed \approach framework focuses on single-line bugs written in the Java programming language, which is the most popular task in previous studies \cite{zhong2022standupnpr}. To evaluate \approach, we first select a large-scale dataset called BFP \cite{tufano2019empirical}, consisting of single-line bugs that can be fixed by using single-line patches within the corresponding buggy methods, as the source to fine-tune the pre-trained model CodeT5. Extensive experiments are then conducted on five benchmarks to validate the performance of \approach. The experimental results demonstrate that \approach can achieve higher accuracy over the state-of-the-art baselines.

The main contributions of this paper are as follows:
\begin{itemize}
  \item We make the first attempt at extracting slicing-based contextual information through program dependence analysis to enhance the NPR task.
  \item We propose an ensemble framework that integrates different NPR models via an effective filter mechanism.
  \item We conduct an extensive evaluation of five commonly used benchmarks to demonstrate the effectiveness of the proposed \approach framework.
\end{itemize}

The remainder of this paper is organized as follows. We describe the related work in Section~\ref{rel}. Section~\ref{met} introduces in detail the proposed framework. We provide the experimental setup in Section~\ref{exp}. Section~\ref{res} shows the analyzing results of our research. We disclose the threats to the validity of our approach in Section~\ref{thr}. Section~\ref{con} draws conclusions and indicates directions for future work.

\section{Related Work}
\label{rel}

\begin{figure*}[t]
  \centering
  \includegraphics[width=\textwidth]{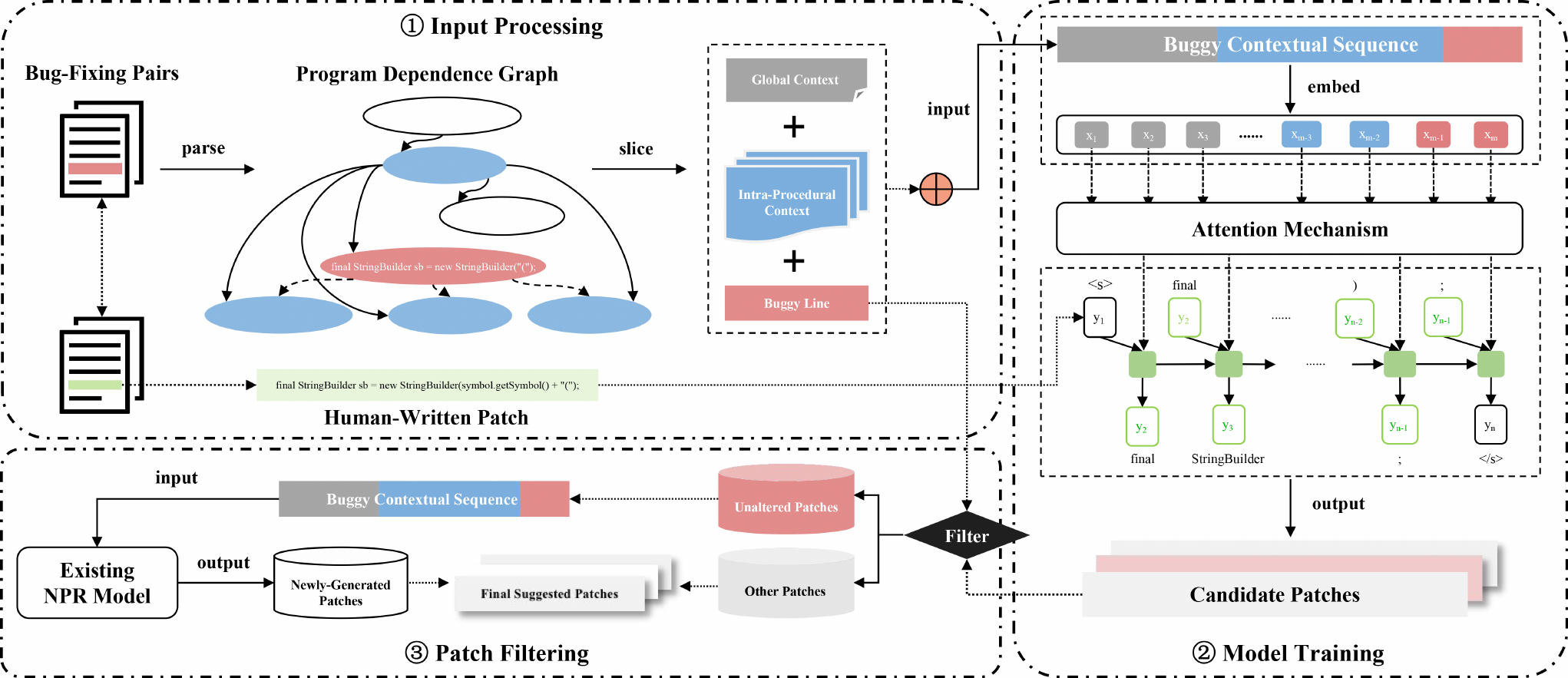}
  \caption{An overview of \approach.}
  \label{overview}
\end{figure*}

As a promising research topic, APR has received significant attention from both the SE and AI communities. According to a living APR review \cite{monperrus2018living}, researchers have proposed a bunch of APR approaches in the last decade, which can be categorized into two mainstreams: search-based \cite{goues2012genprog,qi2014strength,hua2018practical,ghanbari2019practical,liu2019tbar,saha2019harnessing,yuan2020automated,koyuncu2020fixminer} and semantics-based \cite{mechtaev2016angelix,xuan2017nopol,le2017syntax,chen2017contract,afzal2021sosrepair}. With the rapid development of DL-based techniques, researchers have begun to pay more attention to NPR approaches \cite{zhong2022neural,zhang2023survey}, which have demonstrated remarkable potential for improving program repair performance. In contrast to traditional APR approaches, learning-based techniques can automatically capture the semantic relationships between parallel bug-fixing pairs.

Tufano et al. \cite{tufano2019empirical} harness the power of NMT to translate buggy code snippets into fixed ones. In particular, they abstract the identifiers and literals in the source code to reduce the vocabulary size during the data preprocessing process. {\sc SequenceR}\xspace \cite{chen2021sequencer} is an end-to-end approach based on sequence-to-sequence (Seq2Seq) learning that employs the long short-term memory (LSTM) encoder-decoder architecture with copy mechanism to overcome the out-of-vocabulary issue. In addition, {\sc SequenceR}\xspace considers the class-level abstract context surrounding the buggy statement as input to capture the long-range dependencies required for patch generation. Chakraborty et al. \cite{chakraborty2022codit} present a two-stage approach called CODIT to learn code changes for fixing the buggy statements by using an LSTM-based NMT model. DLFix \cite{li2020dlfix} parses the source code to AST and uses a tree-based LSTM to encode the code structures surrounding the bug-fixing changes as contextual information for learning code transformations. Lutellier et al. \cite{lutellier2020coconut} propose an ensemble approach called CoCoNut that combines convolutional neural networks (CNNs) and context-aware NMT models to fix bugs in multiple programming languages. To better capture the diversity of bug-fixing patterns, CoCoNut introduces a novel context-aware NMT architecture that takes the buggy statement and its surrounding method context as two separate inputs. Different from the above NMT- or Seq2Seq learning-based models, Ding et al. \cite{ding2020patching} implement an edit-based model that performs token-level insertion and deletion operations on the buggy code instead of generating raw tokens of the fixed code. Zhu et al. \cite{zhu2021syntax} design Recoder, a syntax-guided edit decoder with a novel provider/decider architecture. Recoder receives the buggy statement with its context, the partial AST, and a tree path from the root node to a non-terminal node as inputs and embeds them using different encoders to generate edits on the AST of the buggy methods.

Inspired by the success of pre-trained models achieved in the field of natural language processing (NLP), an emerging trend is to build language models pre-trained on large source code corpus to boost the performance of program repair. Mastropaolo et al. \cite{mastropaolo2021studying} empirically investigate the performance of the text-to-text transfer transformer (T5) model on four code-related tasks (including program repair). The experimental results indicate that the T5 model can substantially boost the program repair performance. In a follow-up study of CoCoNut \cite{lutellier2020coconut}, Jiang et al. \cite{jiang2021cure} propose CURE that modifies the NMT-based architecture by using a pre-trained GPT module to learn contextual embedding. Mashhadi et al. \cite{mashhadi2021applying} apply a pre-trained model CodeBERT \cite{feng2020codebert} to fix simple Java bugs on the ManySStuBs4J dataset \cite{karampatsis2020single}.

Specifically, \approach utilizes the pre-trained language model to automatically learn bug-fixing patterns from sliced context without necessitating professional domain knowledge. Furthermore, existing NPR models simply consider the context arbitrarily. In contrast, \approach employs the program dependence analysis technique to enhance the NPR task by extracting slicing-based contextual information from PDG. At last, \approach eliminates the unaltered patches via an effective filter mechanism that aims to take advantage of the ensemble performance of different NPR models.

\section{Methodology}
\label{met}

\begin{figure*}[t]
  \centering
  \includegraphics[width=\textwidth]{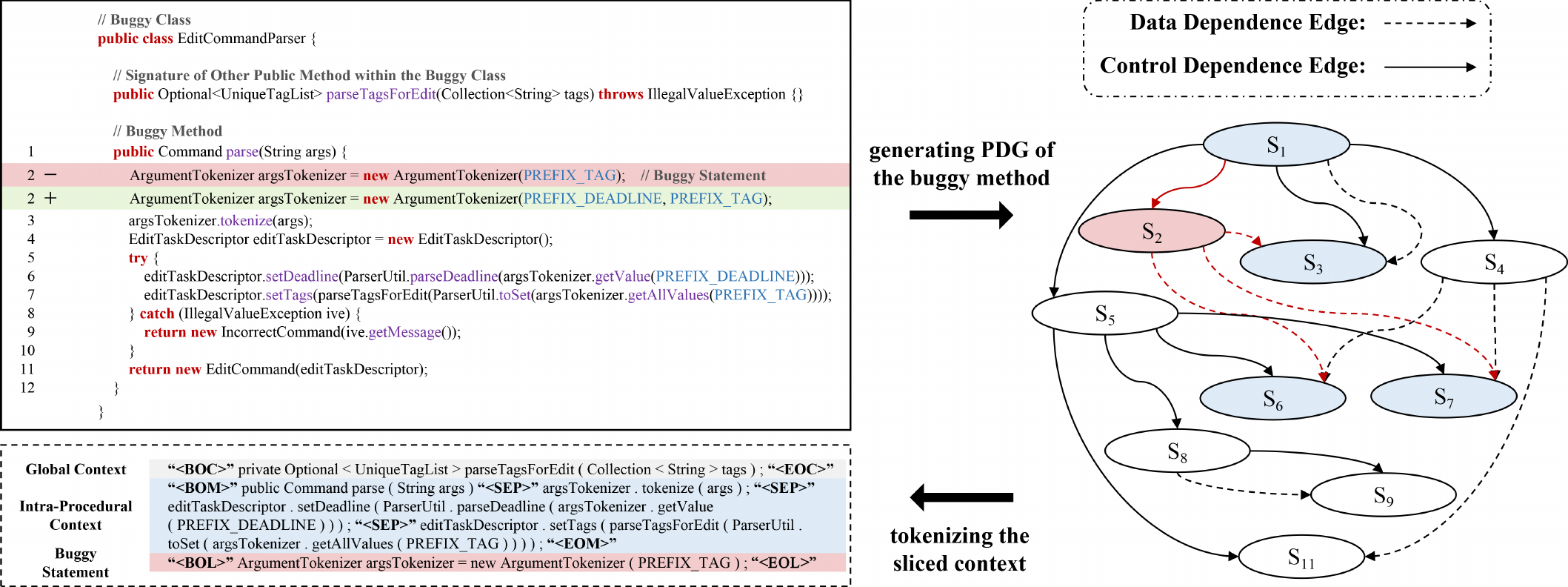}
  \caption{An example input for \approach.}
  \label{input}
\end{figure*}

To address the limitations mentioned in Section~\ref{int}, we present the detailed design (as shown in Figure~\ref{overview}) of our proposed neural APR framework \approach in this section, which mainly consists of three stages: input processing, model training, and patch filtering. We first employ static program slicing on the PDG to extract the relevant contextual information of the corresponding buggy statement. Then, we leverage a general pre-trained language model as the model skeleton to gain the powerful representation learning ability and fine-tune it using the slicing-based contextual inputs. Finally, we integrate different NPR models via an intuitive but effective filter mechanism to filter out the generated unaltered patches.

\subsection{Input Processing}
\label{ip}

Existing NPR approaches \cite{tufano2019empirical,lutellier2020coconut,jiang2021cure,li2022dear,ye2022neural} usually treat the context of a buggy statement in an arbitrary manner, such as by directly considering the whole buggy method or a limited number of lines of code surrounding the buggy statement, which neither captures the semantic relationship between the buggy statement and its context nor includes sufficient contextual information as repair ingredients. Thus, we propose to utilize program slicing techniques to extract contextual information from the graphical representations of source code (i.e., PDGs). Unlike existing approaches, we consider the control and data dependencies of the corresponding buggy statement from the buggy method, which is reported to help capture the bug-fixing patterns \cite{long2016automatic,bohme2017bug}.

\subsubsection{Program Dependence Analysis}

The graphical representations of source code are a fairly used data structure in the field of program analysis \cite{reps1998program}. In this paper, we leverage PDG for program analysis to examine the buggy statement and all the ingredients (e.g., variable usages and method invocations). The PDG explicitly represents the dependencies among statements and predicates, which consists of two subgraphs: the data dependence graph (DDG) and the control dependence graph (CDG). The directed edges in DDG denote a data flow from the source statement to the destination statement, reflecting the influence of one variable on another (i.e., variable definition, usage, or modification). Similarly, each vertex in CDG has a control dependence on its parent vertex. Thus, the control dependency edges reflect the influence of predicates on the values of variables. The upper left corner of Fig.\ref{input} corresponds to a buggy code snippet example, in which Line 2 of the method \texttt{parse} in the \texttt{EditCommandParser} class is a buggy statement that needs to be fixed. To illustrate, the definition of \texttt{argsTokenizer} in statement $S_2$ can lead to its use in statement $S_6$. In this case, a data dependence edge $S_2 \dashrightarrow S_6$ is present in the PDG (shown on the right side of Fig.\ref{input}) of the buggy method.

\subsubsection{Dependency Context Extraction}

To determine all contexts that affect the buggy statement, we introduce an algorithm for extracting dependency context based on the def-use analysis of variables within the given node $\mathrm{N_{buggy}}$. Lines 2 to 19 in the introduced algorithm detailed illustrate the steps of extracting contextual statements by leveraging program slicing (both backward and forward) and ingredient matching. Specifically, we first extract a set of variables that are accessed in $\mathrm{N_{buggy}}$. For each variable $\mathsf{var}$ used in $\mathrm{N_{buggy}}$, we slice a set of nodes $\mathrm{N_{context}}$ from $\mathrm{PDG_{buggy}}$, including both the nodes that have effect on $\mathsf{var}$ and those affected by $\mathsf{var}$. Then, we examine all the ingredients in $\mathrm{N_{context}}$ and store them in $\mathsf{varUsageSet}$ and $\mathsf{methodInvocationSet}$ respectively. In addition, we collect all the public fields $\mathsf{fieldSet}$ and the signatures of other public methods $\mathsf{methodSignatureSet}$ in the buggy class from $\mathrm{AST_{buggy}}$. If the elements in $\mathsf{fieldSet}$ or $\mathsf{methodSignatureSet}$ can match with any of the ingredients in $\mathrm{N_{context}}$, the corresponding AST node of such element is also added to $\mathrm{N_{context}}$ as contextual information. The rationale that motivated this inclusion is that generating a correct patch may require inspecting the status of the used public fields or the argument list of the invoked public methods. Finally, the algorithm outputs the sliced context statements concerning the slicing criterion (i.e., the buggy statement) as repair ingredients to generate the inputs for \approach. In the illustrated example of Fig.\ref{input}, given the buggy statement $S_2$, the algorithm extracts the sliced statements \{$S_1$, $S_3$, $S_6$, $S_7$\} as intra-procedural context and the public method \texttt{parseTagsForEdit} invoked in $S_7$ as global context. The extracted dependency context contains the repair ingredient \texttt{"PREFIX\_DEADLINE"}, thus making it more likely to generate the correct patch for fixing the buggy statement.

\begin{figure}[htbp]
  \centering
  \includegraphics[width=0.47\textwidth]{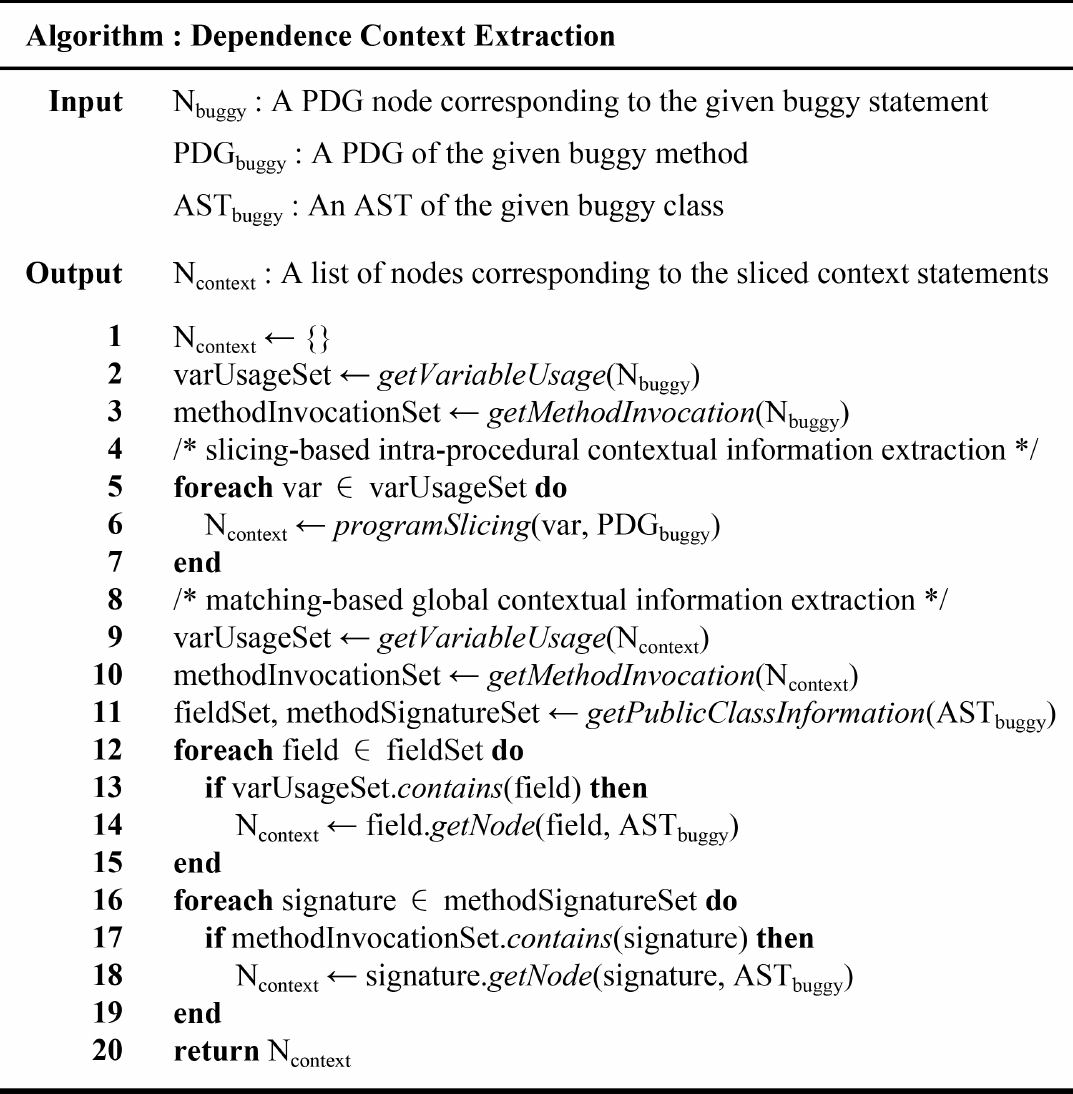}
  \label{algorithm}
\end{figure}

\subsubsection{Context Statement Tokenization}

At this stage, \approach prepares the collected context statements in such a manner that they can be directly fed into the DL models based on an encoder-decoder architecture. The input of \approach is composed of three parts: the buggy statement, the intra-procedural context, and the global context. Existing approaches \cite{lutellier2020coconut,jiang2021cure} attempt to encode different parts of the input separately and then fuse the encoding representations. However, it is still a challenge to effectively eliminate the semantic gaps between different encoders and merge the separated encoding vectors. Recently, the empirical results \cite{chakraborty2021multi} indicate that the best strategy is to encode all of the different input parts using a single encoder. In light of this, \approach employs a consolidated format to combine the buggy statement and its context into one sequence with special isolated tokens. As shown in the lower left corner of Fig.\ref{input}, \approach first separates each content token by a single whitespace. Then, different parts of the input are isolated by different abbreviate tokens (e.g., \texttt{``<BOL>''} denotes the beginning of a buggy statement). \approach also inserts a special token \texttt{``<SEP>''} for each sliced statement in the intra-procedural context to inform the end-of-line information.  In addition, \approach utilizes the sentence-piece tokenizer \cite{kudo2018sentencepiece} to alleviate the open-vocabulary problem \cite{gu2016incorporating}. In this manner, the low-frequency tokens can be synthesized more easily, thus making the program repair task more tractable.

\subsection{Model Training}

As illustrated in the right side of Fig.\ref{overview}, \approach takes the tokenized contextual sequences as input and works on three different pieces of content: 1) the global context that contains information of public fields and method signatures; 2) the intra-procedural context sliced from the buggy method; 3) the buggy statement. We begin with the pre-trained model CodeT5 \cite{wang2021codet5} serving as the reference architecture for the proposed \approach framework. Then, we perform the fine-tuning on the task of program repair. At the inference step, the decoder sequentially predicts the candidate patches.

\subsubsection{Model Architecture}

CodeT5 is a Seq2Seq language model pre-trained on the colossal clean crawled corpus, which consists of an encoder and a decoder. In the context of NPR, an encoder takes a sequence of code tokens as input to map the buggy context code $X=[x_1, \dots ,x_m]$ into a fixed-length hidden state, while the decoder generates the output sequence of tokens $Y=[y_1, \dots ,y_n]$ by taking the hidden state vector as input. Given an input sequence, CodeT5 first obtains the contextualized vector representations for input tokens by projecting them into an embedded vector space through the embedding and positional encoding layer. Then, the vector is fed into the encoder to capture the long-term dependencies from different perspectives of the input sequences. The encoder comprises a stack of Transformer layers, each of which contains a multi-head self-attention layer followed by a position-wise fully connected feed-forward network. The decoder in CodeT5 is a Transformer-based sequential left-to-right decoder that generates one new token at a time until a special stop token is reached. The decoder is similar in structure to the encoder except for the usage of the mask mechanism in multi-head attention, which forces to attend only to past tokens and avoids distraction and information leakage of the subsequent tokens in training.

\subsubsection{Fine-Tuning}

To learn generic bug-fixing representations, we leverage the pre-trained model CodeT5, which has achieved remarkable performance on a variety of NLP and code-related SE tasks, as the starting point to train \approach. We then fine-tune the pre-trained model for the task of program repair. Fine-tuning techniques can optimize the pre-trained parameters to make them more suitable for downstream tasks. Specifically, we represent the program repair task in a ``text-to-text'' format, where the input is the tokenized buggy contextual sequence, and the output is the expected human-written patch. The fine-tuning process is performed using the training corpus of BFP dataset $D$, and each instance within $D$ can be formally represented as a bug-fixing pair $d_i = \{b, f\}$, where $b=(c, m, s)$ comprises the global context obtained from the buggy class $c$, the intra-procedural context sliced from the buggy method $m$ and the buggy statement $s$, and $f$ denotes the human-written patch. The fine-tuning objective is to minimize the cross-entropy loss by learning the mapping $b \rightarrow f$ as a conditional probability $p(f|b)$.

\subsubsection{Patch Inference}

After fine-tuning CodeT5 for the program repair task, \approach can learn latent bug-fixing patterns and generate correct patches. During inference, \approach represents the given buggy statement with three pieces of token sequence: one for the given statement, the other two for its context. \approach follows the typical NPR process and employs beam search to generate the candidate patches sequentially. Once the decoder reaches the stop token, \approach outputs the top-ranked patches for the given buggy statement, where the number of generated candidate patches is configured as beam width.

\subsection{Patch Filtering}

The lower left corner of Fig.\ref{overview} shows the procedure of patch filtering. To avoid the unaltered patching issue existing in the NPR task, \approach proposes an intuitive mechanism to filter out the unaltered patches generated by the pre-trained model CodeT5 for a given bug and then integrates a different NPR model into \approach to re-generate patches for the bug. As discussed in Section~\ref{int}, the unaltered patching issue frequently occurs when using DL-based models for the APR task since the buggy and fixed code snippets are highly similar in the context of fixing single-line bugs. Such an unsupervised phenomenon can be determined by simply comparing the model-generated patches with the input buggy code snippets. In another word, we can directly filter out a portion of the incorrect patches without knowing the human-written ground truth. The goal of patch filtering is to filter out the unaltered patches from the generated candidate patches of the former model and utilize a new NPR model to work on them.

Specifically, given a buggy input, the candidate patches generated by CodeT5 in the first stage can be divided into three categories: correct patches (denoted as $\mathcal{CP\mathtt{_{CodeT5}}}$), unaltered patches (denoted as $\mathcal{UP\mathtt{_{CodeT5}}}$), and other incorrect patches that are inconsistent with the input buggy code snippet (denoted as $\mathcal{IP\mathtt{_{CodeT5}}}$). Then, \approach filters out the $\mathcal{UP\mathtt{_{CodeT5}}}$ and feeds them into another NPR model in the next stage, whereas the $\mathcal{CP\mathtt{_{CodeT5}}}$ and $\mathcal{IP\mathtt{_{CodeT5}}}$ remain as the final suggested patches. At last, the new NPR model will continue to fix the buggy code snippets among $\mathcal{UP\mathtt{_{CodeT5}}}$ with the newly-generated correct patches $\mathcal{CP\mathtt{_{newNPR}}}$. Thus, the correct patches produced by \approach can be formulated as follow:
\begin{equation*}
	\mathcal{CP\mathtt{_{\approach}}} = \mathcal{CP\mathtt{_{CodeT5}}} + \mathcal{UP\mathtt{_{CodeT5}}} \cap \mathcal{CP\mathtt{_{newNPR}}}
\end{equation*}

\section{Experimental Setup}
\label{exp}

\subsection{Experimental Subjects}
\label{subjects}

To evaluate \approach, we require a large-scale corpus of real-world bug-fixing pairs to fine-tune CodeT5. In this paper, we select the BFP dataset \cite{tufano2019empirical} as our original data source, which consists of 787k bug-fixing commits extracted from the GitHub repositories. Each instance is composed of both the buggy and fixed versions of a Java method. We follow the steps described in Section~\ref{ip} to construct our adapted dataset. First, we utilize PROGEX \cite{ghaffarian2021neural}, a cross-platform tool for extracting well-known graphical program representations from source code, to parse each Java method into a corresponding PDG for dependency context extraction. In this step, we filter out the instances that failed to be parsed by PROGEX, meanwhile, we truncate the instances whose length is longer than 512 after subword tokenization. Next, we further split the BFP dataset into training, validation, and testing sets by the ratio of 8:1:1. The dataset split is performed carefully by taking into account possible data leakage. To be specific, any two instances belonging to the same GitHub repository cannot be put in two different sets (e.g., one in training and the other in testing). To enrich the diversity of benchmarks for evaluation, we add four benchmarks widely used in the APR task to our testing set and perform the same procedure of input processing. The detailed statistics for the established benchmarks are listed in Table~\ref{benchmark}. In total, we collect 141195 bug-fixing pairs in the training set, 13523 in the validation set, and 13635 in the testing set. Note that each instance represents a single-line bug that can be fixed by using a single-line patch within one Java method.

\begin{table}[h]\centering
    \caption{Statistics of the five benchmarks.}
    \label{benchmark}
    \begin{tabular}{lccc}
        \toprule
        \textbf{Benchmark} & Training & Validation & Testing \\
        \midrule
        \textbf{BFP} \cite{tufano2019empirical} & 141195 & 13513 & 12224  \\
        \textbf{Bugs.jar} \cite{saha2018bugsjar} & - & - & 1000  \\
        \textbf{Defects4J} \cite{just2014defects4j} & - & - & 260  \\
        \textbf{Bears} \cite{delfim2019bears} & - & - & 119 \\
        \textbf{QuixBugs} \cite{lin2017quixbugs} & - & - & 32 \\
        \bottomrule
    \end{tabular}
\end{table}

\subsection{Experimental Design}
\label{design}

In this section, we introduce the selected baselines, implementation details, and how to assess the generated patches.

\subsubsection{Baseline}
\label{baseline}

To evaluate the performance of our proposed framework, we compare \approach with the following ten baselines that are related to our work:
\begin{itemize}
  \item \textbf{CODIT} \cite{chakraborty2022codit}: a two-stage approach that learns code changes for bug fixing by using a tree-based NMT model.
  \item \textbf{Edits} \cite{ding2020patching}: an edit-based model that performs token-level insertion and deletion operations on buggy code.
  \item \textbf{Tufano} \cite{tufano2019empirical}: an RNN-based model that can translate buggy code snippets into fixed ones.
  \item \textbf{Recoder} \cite{zhu2021syntax}: a syntax-guided decoder to generate edits on the AST of the buggy method.
  \item \textbf{CoCoNut} \cite{lutellier2020coconut}: an ensemble approach for fixing bugs in multiple programming languages that combines CNNs and context-aware NMT models.
  \item \textbf{{\sc \textbf{SequenceR}}\xspace} \cite{chen2021sequencer}: a Seq2Seq learning-based approach for end-to-end program repair that employs the LSTM encoder-decoder architecture with a copy mechanism.
  \item \textbf{RoBERTa} \cite{liu2019roberta}: a Transformer-based model pre-trained on a large corpus of natural language texts in a self-supervised fashion.
  \item \textbf{CodeBERT} \cite{feng2020codebert}: a bimodal pre-trained model for both programming and natural language that learns general-purpose representations to support code-related SE tasks.
  \item \textbf{GraphCodeBERT} \cite{guo2021graphcodebert}: a graph-based pre-trained model for the programming language that considers data-flow information along with code sequences.
  \item \textbf{CodeT5} \cite{wang2021codet5}: a pre-trained encoder-decoder model that better leverages the code semantics conveyed from the developer-assigned identifiers.
\end{itemize}

\footnotetext[2]{https://pytorch.org}
\footnotetext[3]{https://huggingface.co/Salesforce/codet5-small}

\subsubsection{Implementation}
\label{imp}

We implement \approach with the open-source framework PyTorch\footnotemark[2] and initialize \approach with the pre-trained CodeT5-small checkpoint\footnotemark[3] from the Huggingface's website. We adopt the same architecture as T5 \cite{raffel2020exploring} model, consisting of 8-headed attention and 6 layers in both the encoder and decoder. We set the maximum source and target sequence lengths both to 512 and the batch size to 32. For the implementation of the selected baseline models, we use the publicly available source codes provided by the authors or the publicly released checkpoints. During the fine-tuning step, we train \approach for a maximum of 20 epochs. After each epoch, we compute the loss on the validation set and save the model with the minimum validation loss. To avoid over-fitting and save computation costs, we perform early stopping if the validation performance does not improve for five consecutive epochs. During the inference stage, we set both the beam size and candidate number to 10, which means that \approach will generate 10 top-ranked candidate patches for each perfectly fault-localized instance in the testing set for validation. Prior studies usually choose a larger candidate number for evaluation (e.g., Recoder \cite{zhu2021syntax} generates 100 candidate patches). However, recent work \cite{noller2022trust} shows that most developers are only willing to review up to 10 patches. Thus, we also report the evaluation results within the top-10 candidates generated by each baseline in this paper to draw fair conclusions. During the patch filtering step, we integrate {\sc SequenceR}\xspace, which is considered the best NPR model in a recent empirical study \cite{zhong2022standupnpr}, into \approach with our filter mechanism for patch re-generation. We conduct experiments on 4 Nvidia GTX 1080Ti GPUs of 12 GB memory.

\subsubsection{Patch Assessment}

Existing APR approaches typically use test suites for patch validation, which run the human-written test suite against each candidate patch to find plausible patches that can pass all the test cases. Due to the overfitting issue, such plausible patches need further manual assessment to confirm their correctness. However, the empirical results \cite{ye2021automated} indicate that manual patch assessment 1) needs expertise to understand the semantics of the program under repair, 2) may introduce biases to some extent, and 3) can be undoable when the scale is large. To avoid human bias and reduce manual effort, we use a more objective way in our experiment to assess the correctness of each generated candidate patch by checking if it is identical to the human-written one, that is, we use the \textit{exact match} metric to evaluate the model performance on the testing set in this paper.

\section{Results and Analysis}
\label{res}

In this section, we present the experimental results for measuring the performance of our proposed framework and answering the following three research questions (RQs):
\begin{itemize}
  \item \textbf{RQ1}: How does \approach perform compared with the state-of-the-art baselines?
  \item \textbf{RQ2}: How does each component in \approach impact the performance of bug fixing?
  \item \textbf{RQ3}: What is the quality of the candidate patches generated by \approach?
\end{itemize}

\subsection{Answering RQ1}

To answer this question, we compare \approach with ten state-of-the-art baselines on five commonly used benchmarks. To make a fair comparison, we uniformly use the training set of BFP to train or fine-tune the baselines with corresponding input representations via the same training strategy and hyper-parameter settings described in Section~\ref{imp}.

\subsubsection{Experimental Metric Evaluation}

Table~\ref{rq1a} reports the number of correct patches that are identical to the human-written ground truth generated by \approach and the ten baselines. The best result for each benchmark is marked in bold. As shown in Table~\ref{rq1a}, \approach substantially outperforms the ten baselines on all five benchmarks. Specifically, \approach produces more correct patches than the best baseline model CodeT5 by 15.9\% in the BFP benchmark, 12.0\% in the Bugs.jar benchmark, 22.2\% in the Defects4J benchmark, 62.5\% in the Bears benchmark, 7.1\% in the QuixBugs benchmark. As \textit{exact match} is a strict metric, such improvements prove the superiority of \approach in the NPR task. We also observe that the pre-trained models are more promising than those trained from scratch. For instance, CodeT5 improves {\sc SequenceR}\xspace by 14.3\% in the BFP benchmark and 51.5\% in the Bugs.jar benchmark. Since the two models have a similar architecture and a comparable amount of parameters, such improvements demonstrate that using pre-training techniques is advantageous for learning bug-fixing patterns for patch generation. Additionally, the Seq2Seq learning-based models (i.e., \approach, CodeT5, and {\sc SequenceR}\xspace) can obtain better results than other baselines, which illustrates the equal importance of correctly understanding the buggy input and generating the candidate patches.

\begin{table}[h]
\centering
    \caption{Comparison of \approach against the baselines.}
\label{rq1a}
\resizebox{0.47\textwidth}{!}{
    \begin{tabular}{lccccc}
    \toprule
    \multirow{2}{*}{\textbf{Model}} & \textbf{BFP} & \textbf{Bugs.jar} & \textbf{Defects4J} & \textbf{Bears} & \textbf{QuixBugs} \\
    \cmidrule[0.5pt](rl){2-6}
    & {12224 bugs} & {1000 bugs} & {260 bugs} & {119 bugs} & {32 bugs} \\
    \midrule
    \textbf{CODIT} & 44 & 5 & 5 & 1 & 1 \\
    \textbf{Edits} & 117 & 37 & 8 & 2 & 6 \\
    \textbf{Tufano} & 621 & 56 & 18 & 8 & 7 \\
    \textbf{Recoder} & 1003 & 61 & 33 & 1 & 10 \\
    \textbf{CoCoNut} & 1601 & 66 & 37 & 16 & 13 \\
    \textbf{{\sc \textbf{SequenceR}}\xspace} & 1917 & 99 & 38 & 14 & \textbf{15} \\
    \midrule
    \textbf{RoBERTa} & 1373 & 103 & 21 & 8 & 8 \\
    \textbf{CodeBERT} & 1686 & 111 & 29 & 12 & 7 \\
    \textbf{GraphCodeBERT} & 1635 & 115 & 25 & 12 & 6 \\
    \textbf{CodeT5} & 2191 & 150 & 36 & 16 & 14 \\
    \midrule
    \textbf{{\sc \textbf{RepeatNPR}}\xspace} & \textbf{2540} & \textbf{168} & \textbf{44} & \textbf{26} & \textbf{15} \\
    \bottomrule
    \end{tabular}
}
\end{table}

\begin{figure*}[t]
  \centering
  \includegraphics[width=\textwidth]{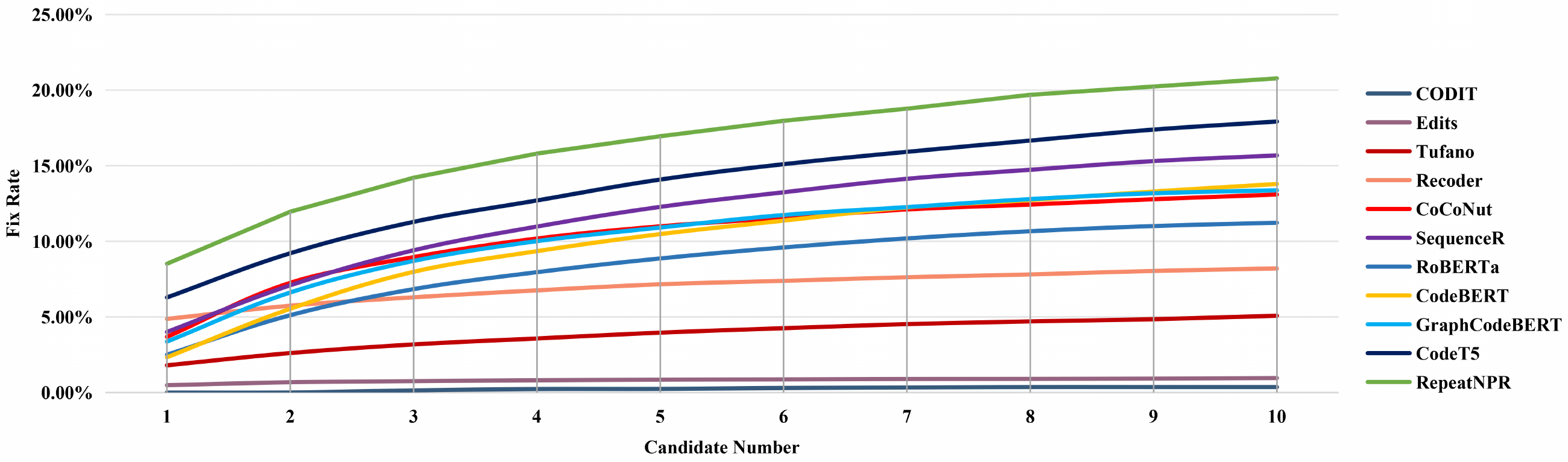}
  \caption{The fixing rates of each NPR model under different candidate numbers.}
  \label{rq1b}
\end{figure*}

\subsubsection{The Impact of Fixing Rate under Different Candidate Numbers}

We further investigate the fixing rates of each corresponding model under different candidate numbers on the BFP benchmark (12224 bugs). We set the candidate numbers from 1 to 10, and the fixing rate denotes the ratio of \textit{exact match} predictions. Note that fixing one bug under candidate number k indicates that at least one of the k model-generated candidate patches is identical to the human-written ground truth. As it can be seen in Fig.\ref{rq1b}, \approach consistently outperforms the ten baselines under all candidate numbers on the BFP benchmark. As the candidate number increases, we notice that the performance gains obtained by different NPR models may be inconsistent, even if they are trained with the unified dataset. For example, when the candidate number is limited to 1, Recoder relatively outperforms {\sc SequenceR}\xspace by 21.4\% in terms of the fixing rate. But when the candidate number reaches 10, the number of correct patches that {\sc SequenceR}\xspace can generate is almost twice that of Recoder.

\subsubsection{Answer to RQ1}

In summary, \approach significantly outperforms the baselines in terms of the \textit{exact match} metric. Such improvements demonstrate the effectiveness of \approach in the APR task. Our observations also indicate that \approach is capable of generating more correct patches under different candidate numbers than the baselines.

\subsection{Answering RQ2}

To answer this question, we evaluate the impact of different components (i.e., program dependence analysis and filter mechanism) in the design of \approach by conducting ablation experiments. For a fair comparison, the training strategy and the hyper-parameter settings are consistent with those described in Section~\ref{imp}.

\subsubsection{Ablation Study}

\begin{figure}[h]
  \centering
  \includegraphics[width=0.47\textwidth]{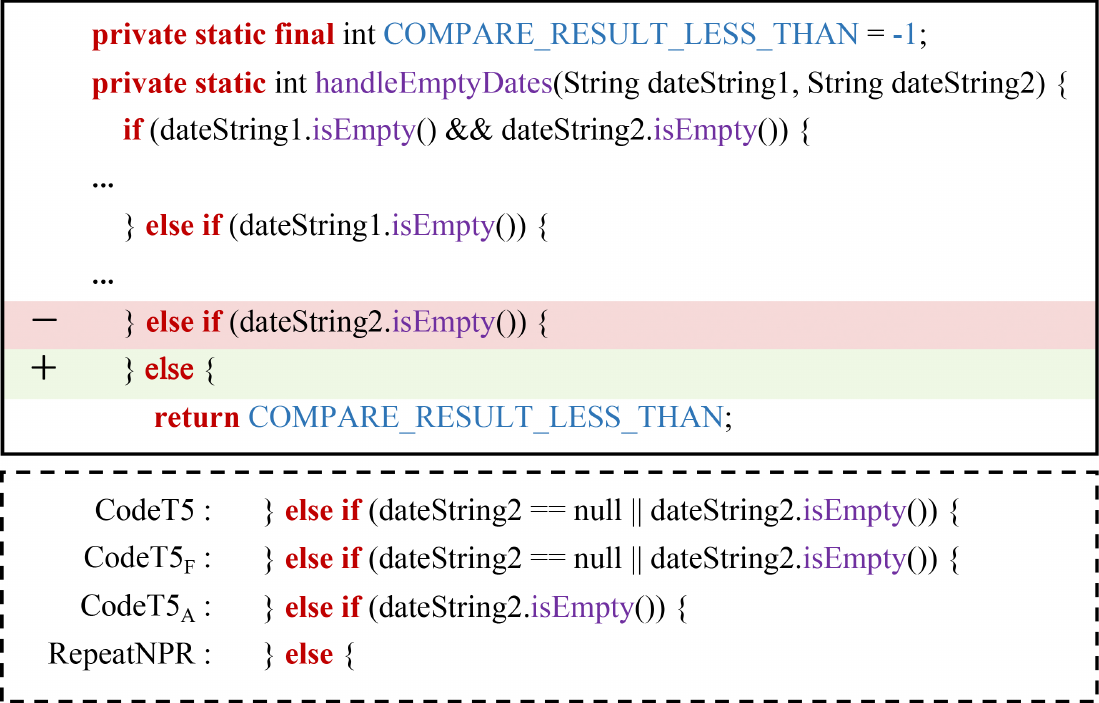}
  \caption{An example from BFP only repaired by \approach.}
  \label{rq2b}
\end{figure}

Table~\ref{rq2a} lists the results with each row representing one model and the number of correct patches that such a model can generate for each benchmark. The best result is marked in bold. To show how each component improves the number of correctly generated patches, we start with the basic pre-trained model CodeT5, which is fine-tuned on the original BFP dataset without using program dependence analysis $\mathcal{A}$ and filter mechanism $\mathcal{F}$. When we augment CodeT5 with component $\mathcal{F}$, the ablated model CodeT5$\mathcal{_{F}}$ respectively repairs 88, 6, 1, 3, and 0 more bugs for the five benchmarks. When we add component $\mathcal{A}$ to CodeT5, the number of correct patches generated by the ablated model CodeT5$\mathcal{_{A}}$ for the five benchmarks is respectively 240, 8, 6, 7, and 1 more than that of CodeT5. Generally, the correct patches generated by the three ablated models are fewer than \approach, which demonstrates the necessity of each component. Figure~\ref{rq2b} depicts a bug-fixing example from BFP only correctly patched by \approach. In this case, the predicate of the buggy statement is redundant according to its context. Thus, a correct patch needs to remove the entire body of the if statement. We can observe that the two ablated models without program dependence analysis (i.e.,  CodeT5 and CodeT5$\mathcal{_F}$) generate the same incorrect patch that is semantically equivalent with the buggy statement, whereas CodeT5$\mathcal{_{A}}$ generates an unaltered patch. Therefore, by further applying component $\mathcal{F}$ on CodeT5$\mathcal{_{A}}$, \approach can successfully produce the correct patch that is identical to the human-written one.

\begin{table}[h]
\centering
    \caption{Ablation study for \approach.}
\label{rq2a}
\resizebox{0.47\textwidth}{!}{
    \begin{tabular}{lccccccc}
    \toprule
    \multirow{2}{*}{\textbf{Model}} & \multicolumn{2}{c}{\textbf{Component}} & \multicolumn{5}{c}{\textbf{Benchmark}} \\
    \cmidrule[0.5pt](rl){2-3}
    \cmidrule[0.5pt](rl){4-8}
    & $\mathcal{A}$ & $\mathcal{F}$ & \textbf{BFP} & \textbf{Bugs.jar} & \textbf{Defects4J} & \textbf{Bears} & \textbf{QuixBugs} \\
    \midrule
    \textbf{CodeT5} & \faTimes & \faTimes & 2191 & 150 & 36 & 16 & 14 \\
    \midrule
    \textbf{CodeT5$\mathcal{_{F}}$} & \faTimes & \faCheck & 2279 & 156 & 37 & 19 & 14 \\
    \midrule
    \textbf{CodeT5$\mathcal{_{A}}$} & \faCheck & \faTimes & 2431 & 158 & 42 & 23 & \textbf{15} \\
    \midrule
    \textbf{{\sc \textbf{RepeatNPR}}\xspace} & \faCheck & \faCheck & \textbf{2540} & \textbf{168} & \textbf{44} & \textbf{26} & \textbf{15} \\
    \bottomrule
    \end{tabular}
}
\end{table}

\subsubsection{The Impact of Program Dependence Analysis on Pre-Trained Models}

\begin{figure}[h]
  \centering
  \includegraphics[width=0.47\textwidth]{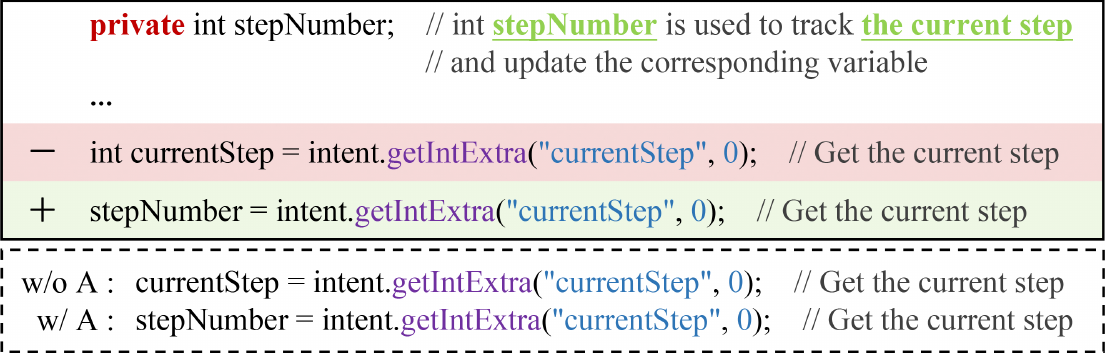}
  \caption{An example from BFP only repaired by CodeT5$\mathcal{_{A}}$.}
  \label{rq2d}
\end{figure}

We further investigate the fixing rate of each pre-trained model when using dependency context as input on the BFP benchmark. Table~\ref{rq2c} presents the comparison results with each pre-trained model comprising two lines of experimental results, in which the first line shows the results of the model that is fine-tuned without using the contextual information based on the program dependence analysis, and the second line shows the results of using such information. The \textbf{Fix@k} metric denotes that a correct patch for a given bug should be among the top-k generated ones. As shown in Table~\ref{rq2c}, we can observe that providing the dependency context as repair ingredients contributes to improving the bug-fixing performance of all the pre-trained models. Among the four pre-trained models, CodeT5 gains the most performance improvements from the input contextual information. This is because CodeT5 is an encoder-decoder model that can leverage the code semantics conveyed from the developer-assigned identifiers to better derive generic representations, whereas the encoder-only pre-trained models (e.g., CodeBERT) treat the source code in the same way as natural language, neglecting the special characteristics of code. Figure~\ref{rq2d} demonstrates a bug that can only be fixed by CodeT5$\mathcal{_{A}}$. As shown in Fig.\ref{rq2d}, the sliced global context provides a hint that variable \texttt{stepNumber} is used to track the current step, thus guiding CodeT5$\mathcal{_{A}}$ to generate the correct patch.

\begin{table}[h]
\centering
    \caption{The fixing rate of with/without dependency context as input for pre-trained models.}
\label{rq2c}
    \begin{tabular}{lrrrr}
    \toprule
    \multicolumn{2}{c}{\textbf{Model}} & \textbf{Fix@1} & \textbf{Fix@5} & \textbf{Fix@10} \\
    \midrule
    \multirow{2}{*}{\textbf{\textbf{RoBERTa}}} & w/o $\mathcal{A}$ & 2.50\% & 8.87\% & 11.23\% \\
    & w/ $\mathcal{A}$ & 2.68\% & 9.60\% & 11.82\% \\
    \midrule
    \multirow{2}{*}{\textbf{\textbf{CodeBERT}}} & w/o $\mathcal{A}$ & 2.33\% & 10.48\% & 13.79\% \\
    & w/ $\mathcal{A}$ & 3.26\% & 10.81\% & 14.24\% \\
    \midrule
    \multirow{2}{*}{\textbf{\textbf{GraphCodeBERT}}} & w/o $\mathcal{A}$ & 3.38\% & 10.91\% & 13.38\% \\
    & w/ $\mathcal{A}$ & 3.39\% & 11.22\% & 13.92\% \\
    \midrule
    \multirow{2}{*}{\textbf{\textbf{CodeT5}}} & w/o $\mathcal{A}$ & 6.29\% & 14.09\% & 17.92\% \\
    & w/ $\mathcal{A}$ & 7.54\% & 15.87\% & 19.89\% \\
    \bottomrule
    \end{tabular}
\end{table}

\subsubsection{The Impact of the Order and Quantity of Models on Filter Mechanism}

We first analyze the impact of models' order on the filter mechanism by comparing $\mathcal{F_{\mathtt{CodeT5\mathcal{_{A}}+SR\mathcal{_{A}}}}}$ and $\mathcal{F_{\mathtt{SR\mathcal{_{A}}+CodeT5\mathcal{_{A}}}}}$ (SR is short for {\sc SequenceR}\xspace). Then, we further integrate a new NPR model GraphCodeBERT$\mathcal{_{A}}$ (short as GCB$\mathcal{_{A}}$) into \approach to evaluate to what extent the models' quantity impacts the performance of the filter mechanism. As it can be seen in Table~\ref{rq2e}, the fixing rates decrease simultaneously when exchanging the models' order, which indicates that the order change may affect the performance of bug fixing. As for the models' quantity, by comparing the results of CodeT5$\mathcal{_{A}}$, $\mathcal{F_{\mathtt{CodeT5\mathcal{_{A}}+SR\mathcal{_{A}}}}}$ and $\mathcal{F_{\mathtt{CodeT5\mathcal{_{A}}+SR\mathcal{_{A}}+GCB\mathcal{_{A}}}}}$, we can see that the fixing rates can be improved or remain the same by integrating new NPR models into \approach. Nevertheless, the performance improvement has decreased with the increase in the models' quantity. A possible reason may be that there is an overlap between the correct patches generated by each model. We will discuss this phenomenon in Section~\ref{overlap}.

\begin{table}[h]
\centering
    \caption{The fixing rate of filter mechanism with different order and quantity of models.}
\label{rq2e}
    \begin{tabular}{lrrr}
    \toprule
    \textbf{Model} & \textbf{Fix@1} & \textbf{Fix@5} & \textbf{Fix@10} \\
    \midrule
    \textbf{GCB}$\mathcal{_{A}}$ & 3.39\% & 11.22\% & 13.92\% \\
    \textbf{SR}$\mathcal{_{A}}$ & 7.58\% & 14.11\% & 17.01\% \\
    \textbf{CodeT5$\mathcal{_{A}}$} & 7.54\% & 15.87\% & 19.89\% \\
    \midrule
    $\mathcal{F_{\mathtt{CodeT5\mathcal{_{A}}+SR\mathcal{_{A}}}}}$ & 8.52\% & 16.95\% & 20.78\% \\
    $\mathcal{F_{\mathtt{SR\mathcal{_{A}}+CodeT5\mathcal{_{A}}}}}$ & 8.14\% & 14.47\% & 17.23\% \\
    $\mathcal{F_{\mathtt{CodeT5\mathcal{_{A}}+SR\mathcal{_{A}}+GCB\mathcal{_{A}}}}}$ & 8.53\% & 16.97\% & 20.80\% \\
    $\mathcal{F_{\mathtt{SR\mathcal{_{A}}+CodeT5\mathcal{_{A}}+GCB\mathcal{_{A}}}}}$ & 8.14\% & 14.47\% & 17.23\% \\
    \bottomrule
    \end{tabular}
\end{table}

\subsubsection{Answer to RQ2}

To sum up, all components of \approach can contribute to performance improvement. Specifically, collecting the dependency context as repair ingredients can better exploit the powerful representation learning ability of the pre-trained model. Meanwhile, the proposed filter mechanism can further increase the number of correct patches by making use of the ensemble performance of different NPR models to eliminate the generated unaltered patches.

\subsection{Answering RQ3}

To answer this question, we manually inspect the generated candidate patches for further evaluation. The evaluation is split into two aspects: 1) analyzing the bug types fixed in the generated correct patches; 2) discussing the overlapping phenomenon among the generated correct patches.

\begin{figure*}[t]
  \centering
  \includegraphics[width=\textwidth]{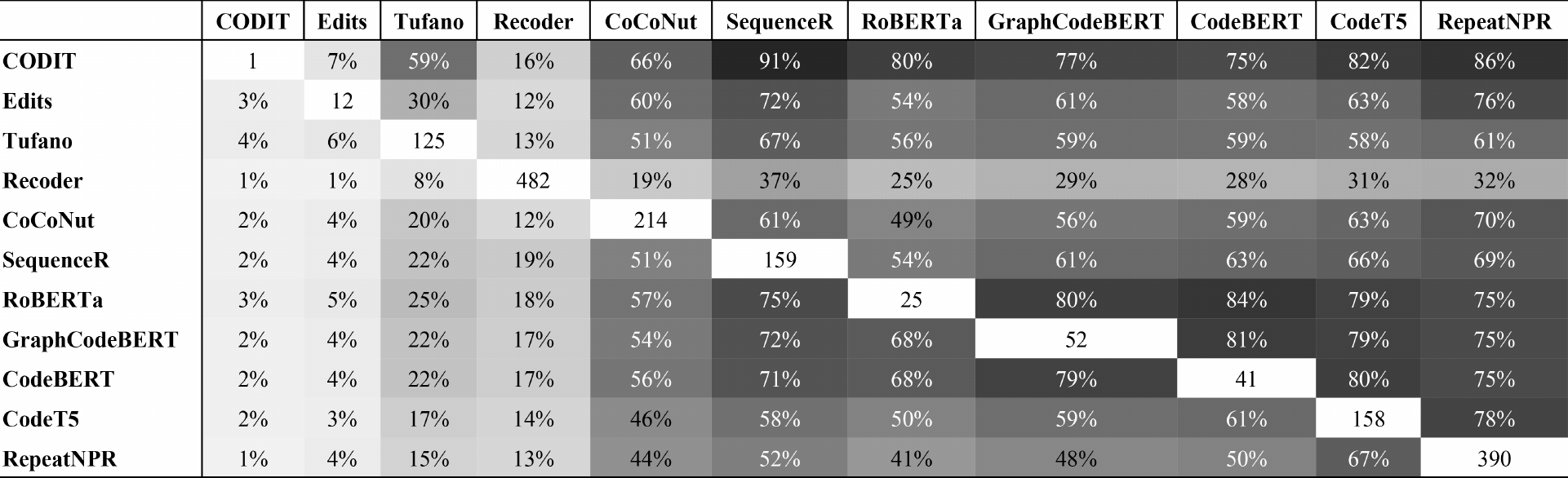}
  \caption{The overlapping patching rates and unique patch numbers of the eleven NPR models.}
  \label{rq3b}
\end{figure*}

\subsubsection{Bug Types Evaluation}

We first compute the difference between two ASTs generated by Spoon \cite{pawlak2016spoon} using the Gumtree algorithm \cite{falleri2014fine}. In total, we classify the bugs into four types, that is Simple Delete, Simple Insert, Simple Replace, and Mixed, according to the edit operations needed to transform one buggy statement into its fixed version. Table~\ref{rq3a} lists the comparison results with each row representing one model and the number of correct patches that such a model can generate for each bug type on the BFP benchmark. The best result is marked in bold. The statistic results shown in Table~\ref{rq3a} indicate that existing NPR models do well in fixing bugs that only need deletion edit operations, but are weak in fixing the more complex ones. That makes sense because the insert and replace edit operations require the model to search for additional tokens to fix the given bug. Specifically, Recoder performs the best on generating code-removal patches, while \approach outperforms all the baselines in fixing the other three types of bugs. Overall, \approach can achieve better or comparable performance than the baseline models in fixing both simple and complex bugs.

\begin{table}[h]
\centering
    \caption{Comparison of the fixed bug types between \approach and the baselines.}
\label{rq3a}
\resizebox{0.47\textwidth}{!}{
    \begin{tabular}{lcccc}
    \toprule
    \multirow{2}{*}{\textbf{Model}} & \textbf{Simple Delete} & \textbf{Simple Insert} & \textbf{Simple Replace} & \textbf{Mixed}\\
    \cmidrule[0.5pt](rl){2-5}
    & {1020 bugs} & {2274 bugs} & {4508 bugs} & {4422 bugs} \\
    \midrule
    \textbf{CODIT} & 2 (0.20\%) & 2 (0.09\%) & 39 (0.87\%) & 1 (0.00\%) \\
    \textbf{Edits} & 16 (1.57\%) & 23 (1.01\%) & 62 (1.38\%) & 16 (0.36\%) \\
    \textbf{Tufano} & 114 (11.18\%) & 121 (5.32\%) & 318 (7.05\%) & 68 (1.54\%) \\
    \textbf{Recoder} & \textbf{503 (49.31\%)} & 73 (3.21\%) & 282 (6.26\%) & 145 (3.28\%) \\
    \textbf{CoCoNut} & 246 (24.12\%) & 399 (17.55\%) & 652 (14.46\%) & 304 (6.87\%) \\
    \textbf{{\sc \textbf{SequenceR}}\xspace} & 391 (38.33\%) & 419 (18.43\%) & 839 (18.61\%) & 268 (6.06\%) \\
    \midrule
    \textbf{RoBERTa} & 279 (27.35\%) & 258 (11.35\%) & 687 (15.24\%) & 149 (3.37\%) \\
    \textbf{CodeBERT} & 283 (27.75\%) & 361 (15.88\%) & 797 (17.68\%) & 245 (5.54\%) \\
    \textbf{GraphCodeBERT} & 302 (29.61\%) & 317 (13.94\%) & 795 (17.64\%) & 221 (5.00\%) \\
    \textbf{CodeT5} & 273 (26.76\%) & 560 (24.63\%) & 1000 (22.18\%) & 358 (8.10\%) \\
    \midrule
    \textbf{{\sc \textbf{RepeatNPR}}\xspace} & 357 (35.00\%) & \textbf{623 (27.40\%)} & \textbf{1104 (24.49\%)} & \textbf{456 (10.31\%)} \\
    \bottomrule
    \end{tabular}
}
\end{table}

\subsubsection{Overlapping Phenomenon Evaluation}
\label{overlap}

As illustrated in Fig.\ref{rq3b}, each row lists the overlapping ratio of correct patches generated for the BFP benchmark between one NPR model and the other models, while the diagonal indicates the number of unique correct patches generated by each model. For instance, 52\% of the correct patches generated by \approach (row 12) can also be fixed by {\sc SequenceR}\xspace (column 7). And there are 390 bugs (row 12, column 12) that can only be fixed by \approach. According to the results in Fig.\ref{rq3b}, we observed that the better the repair performance of a model, the higher the overlapping patching rate between it and other models. Regarding the results in Table~\ref{rq1a}, we can find that our proposed framework \approach, CodeT5, and {\sc SequenceR}\xspace are the top three models evaluated on the BFP benchmark. Relatively, the overlapping rate of other models with the three ones is much higher. The possible reason for this phenomenon may be that the DL-based approaches mostly adopt similar network architectures and inference paradigms. Furthermore, the number of unique correct patches generated by \approach is larger than that of other baselines (except Recoder). As is discussed in Section~\ref{rel}, the possible reason is that Recoder is designed to generate edits on the AST of the buggy methods, which is different from the NMT- or Seq2Seq learning-based approaches. Nevertheless, regarding the results we obtained in Table~\ref{rq3a}, Recoder tends to fix bugs of a specific type (i.e., Simple Delete).

\section{Threats to Validity}
\label{thr}

In this section, we illustrate the main threats to the validity of our approach, which are listed as follows:
\begin{itemize}
	\item \textbf{External threat}: The quality of the selected experimental subjects and the generalizability of \approach are the principal threats to external validity in this paper. First, we use the mainstream dataset BFP for fine-tuning as previous studies \cite{tufano2019empirical,tang2021grammar,wang2021codet5,zhong2022standupnpr}. We remove all duplicate instances between the training and testing sets to avoid the data leakage issue. Second, \approach has been evaluated in Java bugs. Besides, the designed components in \approach are language-agnostic and can be applied to other programming languages.
	\item \textbf{Internal threat}: It is widely known that DL-based models are sensitive to hyper-parameters. Thus, using a sub-optimal hyper-parameter can pose an internal threat to the validity of \approach. Due to the limitation of computational resources, we cannot thoroughly explore optimal hyper-parameters in this paper. Since Raffel et al. \cite{raffel2020exploring} have explored effective settings of hyper-parameters through extensive experiments in previous work, we use the same hyper-parameters described in their paper. We acknowledge that there might be room for further improvement through additional tuning.
	\item \textbf{Construct threat}: In this paper, the experimental metric used for model evaluation is referred to as the construct threat. We adopt the \textit{exact match} metric to assess the correctness of the generated candidate patches. Although such a metric does not represent human judgment, it is a strict and objective metric that can be used to quickly and quantitatively evaluate the model performance. In the future, we will conduct more human evaluations.
\end{itemize}

\section{Conclusion and Future Work}
\label{con}

In this paper, we propose a novel NPR framework called \approach that adapts a state-of-the-art encoder-decoder language model for fixing single-line Java bugs. To accurately capture the semantic relationship between the buggy statement and its context, we make the first attempt to leverage program dependence analysis to improve the NPR task by extracting slicing-based contextual information as repair ingredients from PDG. Additionally, we propose an intuitive but effective filter mechanism to eliminate the unaltered patches by taking advantage of the ensemble performance of different NPR models. Empirical results demonstrate that \approach outperforms the state-of-the-art approaches on five widely used benchmarks in terms of the \textit{exact match} metric. In the future, we plan to expand the program analysis techniques to collect project-specific knowledge for bug fixing. Since the PDG and AST exist in most programming languages, we can extend \approach with small modifications to support new target languages. Furthermore, we will design a better filter mechanism that can determine more incorrect patches (except the unaltered ones) without comparing them with the human-written ground truth.

\bibliographystyle{IEEEtran}
\bibliography{IEEEabrv,npr}

\end{document}